Abschlussbericht

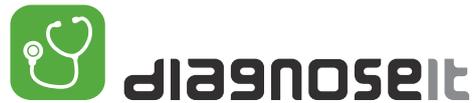

# Expertengestützte automatische Diagnose von Performance-Problemen in Enterprise-Anwendungen

Christoph Heger,[1] André van Hoorn,[2] Dušan Okanović,[1]
Stefan Siegl,[1] Christian Vögele,[1] Alexander Wert[1]

[1] NovaTec Consulting GmbH, Leinfelden-Echterdingen
[2] Universität Stuttgart, Inst. für Softwaretechnologie, Stuttgart

# Inhaltsverzeichnis

# 1 Einleitung

Nichtfunktionale Qualitätseigenschaften von geschäftskritischen, betrieblichen Anwendungssystemen (im Folgenden als Enterprise-Anwendungen bezeichnet) haben einen wesentlichen Einfluss auf die betriebswirtschaftlichen Kenngrößen eines Unternehmens. Eine der wichtigsten Qualitätseigenschaften ist Performance, welche den Grad quantifiziert, zu dem eine Anwendung die Anforderungen in Bezug auf das Zeitverhalten und die Ressourcenauslastung einhält. So wirken sich beispielsweise Performance-Probleme wie zu hohe Antwortzeiten negativ auf die Kundenzufriedenheit und damit auf den Unternehmensumsatz aus. Zur Überwachung der Performance von Enterprise-Anwendungen im Betrieb werden Prozesse und Werkzeuge zum Application Performance Management (APM) in den Application Lifecycle integriert. Diese stellen sicher, dass Performanceprobleme möglichst frühzeitig gefunden und bereinigt werden. Primär werden APM-Werkzeuge zur Erkennung und Signalisierung von Performance-Problemen bzw. Symptomen (kontinuierliches Monitoring) und zur Ermittlung der Fehlerursache (Diagnose) verwendet. Im nachgelagerten Schritt erfolgt die Auflösung des Problems. Für die Qualität des Monitorings und der Diagnose ist die Qualität der Messdaten essentiell. Die Qualität hängt stark von der Messdatentiefe und -qualität, d. h. z. B. welche Daten in welcher Häufigkeit erfasst werden, ab. Problemfälle können im Monitoring nur erkannt werden, wenn entsprechende Messdaten vorhanden sind. Eine detaillierte Diagnose erfordert eine noch tiefere Messdatensammlung, um die Ursache des Problems ermitteln zu können.

Erfahrungen aus der APM-Praxis bei großen Enterprise-Anwendungen zeigen, dass es mit den aktuell gängigen APM-Prozessen und -Werkzeugen insbesondere folgende Probleme gibt: *i)* aufwändige initiale Konfiguration (Instrumentierung) und deren Aktualisierung, *ii)* manuelle Problemerkennung und -diagnose (vom Symptom zur Ursache), *iii)* jeder APM-Prozess ist für sich losgelöst. Diese werden im Folgenden detailliert.

*Aufwändige und ständige Aktualisierung von APM-Konfigurationen.* Die Konfiguration von APM-Prozessen und -Werkzeugen ist kompliziert und zeitintensiv. Bei aktuellen APM-Lösungen muss der Endbenutzer eigenständig die Instrumentierung in Form von Messpunkten (z. B. zur Messung von Antwortzeiten bestimmter Methoden) für das zu analysierende System ermitteln und das APM-Werkzeug entsprechend konfigurieren. Die Herausforderung besteht im Trade-Off zwischen Informationsgewinnung und dem resultierenden Overhead. Die Erstellung der Konfiguration ist immer ein iterativer Prozess, für welchen neben technischem Verständnis und Erfahrung in der Fachdomäne auch Erfahrungen im Bereich APM — zur Definition der Messpunkte im konkreten Werkzeug — notwendig sind. Da die Ermittlung einer sinnvollen Konfiguration für eine Anwendung häufig nur durch Trial-and-Error in einer produktionsähnlichen Testumgebung mit produktionsähnlichem Nutzungsprofil möglich ist — und viele APM-Werkzeuge einen Neustart des Systems erwarten — ist die Erstellung von guten Konfigurationen sehr zeitintensiv. Software-Evolution bedingt zudem, dass die APM-Konfiguration stetig angepasst werden muss. Durch das vermehrte Einsetzen von agilen Entwicklungsmethodiken und die damit verbundenen häufigen Auslieferungen ist somit ein hoher Aufwand für die Erstellung und Wartung der Konfiguration notwendig, die von bisherigen Werkzeugen nicht unterstützt wird. In vielen Projekten ist deswegen die Konfiguration nicht optimal und beeinflusst die Qualität der kompletten APM-Lösung stark negativ.

*Manuelle Problemerkennung und -diagnose.* Die Erkennung und Diagnose von Performance-Problemen erfolgen manuell. Die zur Laufzeit erhaltenen Messungen müssen im einfachsten Fall mit Schwellwerten belegt werden, so dass im Falle eines Problems, z. B. hohe Antwortzeiten, eine Alarmierung ausgesendet werden kann. Die Definition sinnvoller Schwellwerte gestaltet sich in der Praxis schwierig, da diese Schwellwerte häufig selbst beim Anwendungsseigner und im Fachbereich nicht bekannt sind und sich häufig ändern können. Oftmals müssen daher die Schwellwerte während des Betriebs der Anwendung gelernt und im Werkzeug integriert werden. Im Fehlerfall muss ein APM-Experte die Messpunkte analysieren und das Problem auf Basis der Symptome und der Messwerte diagnostizieren. Jedes Symptom kann durch zahlreiche Ursachen ausgelöst werden. Während der Diagnose muss der Analyst daher systematisch die möglichen Ursachen prüfen und die konkrete Ursache ermitteln. Ohne technische Fachkenntnis und Erfahrungsschatz im Bereich Performance-Diagnose werden jedoch häufig nicht alle möglichen Ursachen untersucht und dadurch Performance-Probleme nicht gelöst bzw. sogar manchmal verstärkt. Auch kommt es häufig vor, dass nicht genügend Messdaten vorliegen. Um diese Situation zu erkennen, ist vor allem ein hohes Maß an APM-Erfahrung erforderlich. Neben der technischen Komplexität ist dieser Prozess zudem sehr zeitintensiv. Da in Projekten APM-Experten — die zudem noch fachliches Verständnis der Anwendungen aufweisen — Mangelware sind, werden Diagnosen häufig nicht erfolgreich durchgeführt.

*Losgelöstheit von APM-Prozessen.* Zur erfolgreichen Anwendung von APM ist somit aktuell ein APM-Experte erforderlich, der für jede Anwendung eine optimale Konfiguration erstellt und stets aktuell hält sowie Performance-Probleme erkennt und diagnostiziert. In großen Unternehmen sind häufig mehrere APM-Experten notwendig, um die Anwendungen sinnvoll unterstützen zu können. Jeder Experte arbeitet jedoch meist isoliert basierend auf seinem Fachwissen. Aktuelle Werkzeuge bieten nur bedingte Möglichkeiten zum Wissensaustausch zwischen APM-Experten, so dass APM-Erfahrungen weder für eine konkrete Anwendung noch anwendungsübergreifend systematisch geteilt werden können.

Das Ziel des Projekts *diagnoseIT* war es, die genannten Problemstellungen zu adressieren. Die genaue Aufgabenstellung wird im folgenden Kapitel erläutert. Es folgt eine Übersicht über die erzielten Ergebnisse und Erfahrungen.

## 2 Kurzdarstellung

### 2.1 Aufgabenstellung

Ziel des *diagnoseIT* Projekts ist es, aufbauend auf formalisiertem APM-Expertenwissen bestehende APM-Prozesse um eine automatische Konfiguration und Kontrolle der Messpunkte, sowie eine automatische Problemerkennung und -diagnose zu erweitern. Somit richtet *diagnoseIT* den APM-Prozess und die genutzten Werkzeuge optimal auf die Bedürfnisse der gegebenen Enterprise-Anwendung aus, um den Automatisierungsgrad in APM-Prozessen zu erhöhen.

*diagnoseIT* übernimmt selbstständig die Steuerung der Messpunkte und kommuniziert dies an die APM-Lösung. Dadurch wird — ohne menschliches Zutun — ein optimales Abwiegen zwischen Detaillierungsgrad und Overhead erreicht. Die Instrumentierung startet hierbei initial durch eine Out-of-the-Box-Instrumentierung auf Basis von Expertenwissen. Diese Instrumentierung wird automatisch verfeinert und somit an die Bedürfnisse der Anwendung angepasst. Eine automatische Problemerkennung löst die fehleranfällige und schwer wartbare manuelle Definition von Schwellwerten ab. Zudem kennt die *diagnoseIT*-Lösung bereits gängige Fehlermuster und führt auf Basis der erkannten Symptome eine automatische Diagnose durch. Beides erfolgt durch Einbeziehung des erfassten APM-Expertenwissens. Sollte die Detaillierungstufe der Metriken für die Problemerkennung und/oder die Fehlerdiagnose zu gering sein, kann das System selbstständig die Detaillierung durch Erweiterung der Datensammlung verbessern. Werden die Informationen nicht mehr benötigt, werden die Datenpunkte deaktiviert, um den Overhead zu minimieren.

## 2.2 Voraussetzungen

Die wesentlichen technischen Voraussetzungen zur Durchführung des Projekts wurde durch eine Vielzahl an APM-Werkzeugen erfüllt. Zudem sind besonders das an der Universität Kiel entwickelte Kieker-Framework zur dynamischen Analyse von Softwaresystemen oder das bei NovaTec Consulting GmbH entwickelte APM-Werkzeug inspectIT zu nennen. Basierend auf diesen Werkzeugen werden Traces und andere Laufzeitdaten erhoben. Sowohl die NovaTec als auch die Universität Stuttgart verfügten über umfassende Kenntnisse in den Bereichen APM bzw. Performance-Engineering.

Eine weitere wichtige organisatorische Voraussetzung war die Verfügbarkeit von Fallstudien, an denen die Strategien, Techniken und Werkzeuge erprobt werden konnten. Diese Fallstudien wurden von den zwei assoziierten Partnern bereitgestellt, die das Projekt begleiteten.

## 2.3 Planung und Ablauf

Die Laufzeit des Projekts *diagnoseIT* erstreckte sich von März 2015 bis Juni 2017.

Das Projekt wurde in die folgenden sechs Arbeitspakete (AP 1–6) strukturiert. Eine grafische Darstellung der inhaltlichen Arbeitspakete AP 1–4 und deren Beziehungen ist in Abb. 1 (Seite 6) enthalten.

- *AP 1 (Enterprise-Performance-Modell).* Das Arbeitspaket umfasst sämtliche Aktivitäten zur Repräsentation der Informationen über das überwachte Enterprise-System.

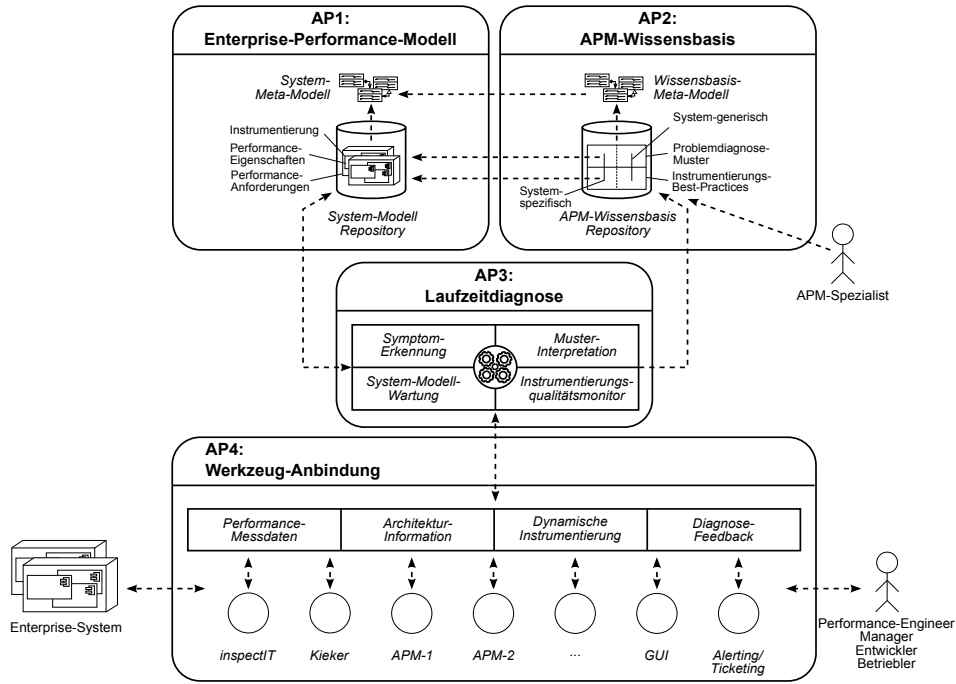

Abbildung 1: Übersicht über den *diagnoseIT*-Ansatz und die inhaltlichen Arbeitspakete

- *AP 2 (APM-Wissensbasis).* Das Arbeitspaket umfasst sämtliche Aktivitäten zur Erstellung einer APM-Wissensbasis sowie deren Befüllung mit APM-Expertenwissen.

- *AP 3 (Laufzeitdiagnose)* Das Arbeitspaket umfasst sämtliche Aktivitäten zur Durchführung der Laufzeitanalysen auf Basis der Systeminformationen (AP 1 und AP 4) sowie der APM-Wissensbasis (AP 2). Zu den Analysen zählen die Extraktion von Architekturinformation, die Erkennung und Diagnose von Performance-Problemen sowie Anpassung der Instrumentierung.

- *AP 4 (Werkzeug-Anbindung).* Das Arbeitspaket umfasst sämtliche Aktivitäten zur Entwicklung der Schnittstelle zu den APM-Werkzeugen. Ebenfalls Teil dieses Arbeitspaketes ist die Integration und Erweiterung der Werkzeuge inspectIT und Kieker.

- *AP 5 (Evaluation von* diagnoseIT*).* Das Arbeitspaket umfasst sämtliche Aktivitäten zur qualitativen und quantitativen Bewertung der in den Arbeitspaketen AP 1–4 erarbeiteten Ergebnisse. Die Evaluation erfolgt in Form von Labor- und Fallstudien.

- *AP 6 (Projektkoordination und Dissemination).* Das Arbeitspaket umfasst sämtliche Aktivitäten, die mit der Organisation, Erfolgskontrolle, Steuerung und Berichterstattung (Zwischenberichte und Abschlussbericht) des Projektes verbunden sind sowie die in *diagnoseIT* erarbeiteten Forschungsergebnisse während der Projektdurchführung einer breiteren Öffentlichkeit zur Verfügung zu stellen.

An allen Arbeitspaketen waren die beiden geförderten Projektpartner (NovaTec und Universität Stuttgart) beteiligt. Darüberhinaus erfolgte eine Einbindung der assoziierten Partner in AP 1, AP 2, AP 4 und AP 5.

### 2.4 Wissenschaftlicher und technischer Stand vor Projektdurchführung

Die zu *diagnoseIT* verwandten Themengebiete aus Wissenschaft und Technik, auf die im Folgenden eingegangen wird, sind Sprachen und Technologien zur Modellierung von Softwaresystemen, modell- und messbasierte Performance-Analyse sowie APM-Werkzeuge.

**Sprachen und Technologien zur Modellierung von Softwaresystemen** Zahlreiche Modellierungssprachen zur Beschreibung von Softwaresystemen auf verschiedenen Abstraktionsebenen wurden in den vergangenen Jahren sowohl im Bereich der Forschung als auch in der industriellen Praxis entwickelt [TMD09]. Beispiele für Architekturbeschreibungssprachen (ADLs) sind AADL, ACME und xADL. Die Unified Modeling Language (UML) der Object Management Group (OMG) erlaubt die Beschreibung von objektorientierten Softwaresystemen. Weitere Modellierungssprachen der OMG zur Systembeschreibung sind KDM und ASTM. Letztere erlaubt die Systembeschreibung auf Basis von abstrakten Syntaxbäumen.

Darüberhinaus stehen Modellierungssprachen bereit, um Systemmodelle mit Informationen zu nichtfunktionalen Eigenschaften anzureichern, beispielsweise um Performance-Anforderungen oder Eigenschaften auszudrücken. Beispiele für diese Sprachen sind die UML-Profile SPT und MARTE , die ebenfalls Spezifikationen der OMG sind.[1] Heutzutage stehen ausgereifte, in der Wissenschaft und Praxis breit verwendete, Technologien zur modellbasierten und modellgetriebenen Entwicklung (MDE) [SV06, BCW12] in Form von Meta-Modellierungs- und Transformationssprachen zur Verfügung; beispielsweise aus dem Eclipse Modeling Project (EMP) [St09].

**Modell- und messbasierte Performance-Analyse** Ansätze zur Performance-Analyse lassen sich in modellbasierte und messbasierte Ansätze unterscheiden [Li05]. Ziel modellbasierter Ansätze [CDMI11], beispielsweise Palladio [BKR09], ist es, die Performance eines Systems vorherzusagen, bevor eine Implementierung des Systems existiert. Häufig dienen oben genannte Systemmodelle, die um nichtfunktionale Eigenschaften angereichert sind, als Basis zur Analyse. Bei messbasierten Ansätzen werden Messdaten eines laufenden System aufgezeichnet und analysiert. Dies kann sowohl in Test- als auch in Produktionsumgebungen geschehen. Zu beachten ist, dass messbasierte Ansätze immer einen Performance-Einfluss auf das beobachtete System haben.

---
[1] http://www.omg.org/spec

Sowohl auf modellbasierter als auch messbasierter Ebene wurden verschiedene Techniken und Ansätze zur automatischen Erkennung, Diagnose und Auflösung von Performance-Problemen entwickelt. Grundlage zur Erkennung und Diagnose von Performance-Problemen sind typischerweise bekannte statistische Techniken wie Zeitreihenanalyse, Trenderkennung und maschinelles Lernen, ggf. unter Hinzunahme von Architekturinformation [Ma09]. Adaptive Ansätze zur Problemdiagnose [Mo04, Eh11] schalten Messpunkte dynamisch zu und wieder ab, beispielsweise um den Einfluss auf das System zu reduzieren. Es ist geplant, existierende Ansätze in *diagnoseIT* zu integrieren und auszubauen. Einige Ansätze zur Erkennung und Diagnose basieren auf dokumentierten Performance-Anti-Pattern und beziehen damit APM-Expertenwissen mit ein [WHH13, Tr14]. Sie fokussieren jedoch nicht auf die automatische Erkennung und Diagnose in Produktion.

**APM-Werkzeuge**  Die Gartner APM-Studie von 2013 ([KC13]) umfasst 13 kommerzielle APM-Werkzeuge; vier davon (AppDynamics, CompuwareAPM, NewRelic und Riverbed Technology) werden von Gartner als Leader im Bereich APM geführt. Es sind nur wenige freie oder Open-Source-Lösungen vorhanden, welche neben Monitoring eine tiefgreifende Diagnose ermöglichen. Die Antragsteller NovaTec und Universität Stuttgart entwickeln die freien APM-Werkzeuge inspectIT und Kieker. Die meisten APM-Werkzeuge verwenden agentenbasierte Lösungen, die zur Laufzeit Messpunkte in den Bytecode der Anwendung integrieren. Neben instrumentierenden, agentenbasierten Lösungen bietet der Markt auch agentenbasierte Ansätze, die auf Stacktrace-Sampling beruhen. Diese Lösung, vor allem getrieben durch das Werkzeug AppDynamics, kann detaillierte Informationen in Ablaufgraphen aufzeigen, ohne dass eine Instrumentierung notwendig ist. Verglichen mit einem instrumentierenden Ansatz verursacht das Stacktrace-Sampling jedoch einen deutlich höheren Overhead. Deswegen limitieren Sampling-basierte Lösungen die Anzahl der Messdaten, wodurch die Sichtbarkeit leidet. Viele instrumentierende Ansätze können aufgrund des geringeren Overheads für jede Anfrage einen detaillierten transaktionalen Aufrufbaum schreiben. Sampling-basierte Ansätze sind einfacher zu konfigurieren, erkaufen sich dies mit reduzierter Sichtbarkeit.

APM-Lösungen bieten derzeit keine hinreichende Lösung zur adaptiven Anpassung der Instrumentierungstiefe zur Laufzeit. Typischerweise werden lediglich Konfigurationen für unterschiedliche Technologien als Out-of-the-Box-Konfiguration angeboten. Diese Konfigurationen sind jedoch statisch und werden nicht durch das Werkzeug zur Laufzeit verfeinert. Auf Basis der Out-of-the-Box-Konfiguration bieten einige Hersteller Out-of-the-Box-Monitoringansichten an (z. B. Compuware und AppDynamics). Typischerweise handelt es sich hierbei jedoch lediglich um Prüfungen der Endbenutzerantwortzeit und Fehlerraten, gemessen ab Einstieg in die Anwendung. Verglichen werden diese Zeiten gegen manuell definierte Schwellwerte oder — in manchen Werkzeugen — gegen eine vom

Werkzeug erstellte Baseline. CA Wily Introscope bietet seit wenigen Monaten mit *Prelert* eine automatisierte Anomalieerkennung, um Probleme in Anwendungen basierend auf den gesammelten Messdaten zu ermöglichen. Eine automatisierte Diagnose wird jedoch von bisher keinem kommerziellen Werkzeug angeboten. In jedem Werkzeug muss die Diagnose durch einen Performance-Analysten durchgeführt werden. Automatisierte Unterstützung ist nicht vorhanden.

### 2.5 Zusammenarbeit mit anderen Stellen

Während der Projektlaufzeit wurde intensiv mit den assoziierten Partnern zusammengearbeitet. Ferner erfolgte eine Zusammenarbeit mit weiteren wissenschaftlichen Partnern wie der Research Group der Standard Performance Evaluation Corporation (SPEC) sowie den Kollegen Alberto Avritzer (USA) und Catia Trubiani (Italien). Hinzu kommen Interaktionen mit Herstellern kommerzieller Werkzeuge.

## 3 Eingehende Darstellung

Im Folgenden werden die im Projekt *diagnoseIT* erzielten Ergebnisse vorgestellt. Die Darstellung der Ergebnisse orientiert sich an den in Abschnitt 2.3 genannten Arbeitspaketen und werden im Folgenden beschrieben.

### 3.1 AP 1: Enterprise-Performance-Modell

Ziel des Arbeitspakets war die Entwicklung relevanter Repräsentationen der Informationen über das überwachte Enterprise-System.

Begonnen wurde mit der Enwicklung eines *diagnoseIT*-Meta-Modells zur Repräsentation eines sogenannten Enterprise-Performance-Modells (EPM) sowie des Modell-Repositories. Das *diagnoseIT*-Meta-Modell erlaubt es, die für den diagnoseIT-Ansatz benötigten Informationen über das überwachte Enterprise-System zu repräsentieren. Das Modell-Repository bietet eine Schnittstelle zum Zugriff auf das Systemmodell mittels entsprechender Abfragen (Queries). Eine erste Version des EPM wurde definiert und mittels des Eclipse-Modelling Frameworks (EMF) als Ecore-Meta-Modell umgesetzt. Das EPM enthält Konzepte zur Modellierung der Software-Entitäten (inkl. Annotationen wie verwendete Rahmenwerkzeuge und O/R-Mapper) sowie deren Verteilung auf die Laufzeitumgebung (Server inkl. Hardware-Ressourcen, virtuelle Maschinen, etc.). Es wurde ein Modell-Repository auf Basis von CDO (Common Data Objects) umgesetzt, welches einen datenbankähnlichen Zugriff auf EMF-Modelle erlaubt.

Eine frühe Entscheidung während des Projektverlaufs war es, dass wir nach Erstellung des EPM, welches eine zustandsbasierte Analyse erlaubt, im Folgenden anstreben, einen Ansatz zu entwickeln und zu evaluieren, der möglichst zustandslos ist (Diagnose allein durch die Analyse einzelner Traces). Dadurch wurde eine deutliche Steigerung von Durchsatz und Skalierbarkeit angestrebt.

Für den Austausch von Execution Traces zwischen APM-Werkzeugen wurde ein initiales Datenmodell in Form einer Programmierschnittstelle namens OPEN.xtrace (vormals als Common Trace API bezeichnet) entwickelt. Hinzu kommen Adapter für die APM-Werkzeuge inspectIT, Kieker, Dynatrace, AppDynamics und CA Introscope. Der Ansatz wurde u. a. mit den assoziierten Partnern und mit führenden APM-Herstellern diskutiert. Aus den Diskussionen sind weitere Anforderungen bzgl. der Ausdrucksmächtigkeit gesammelt worden. Es wurde eine Studie mit führenden APM-Werkzeugen (kommerziell und Open-Source) in Bezug auf die für Traces gelieferten Daten durchgeführt. Die Studie beleuchtete insbesondere zwei Fragen: i) Welche von OPEN.xtrace unterstützten Informationen werden von den APM-Werkzeugen (nicht) unterstützt? ii) Welche der von den APM-Werkzeugen gelieferten Informationen werden von OPEN.xtrace aktuell nicht unterstützt? Die Studie diente u. a. dazu, OPEN.xtrace um zusätzliche Ausdrucksmächtigkeit zu erweitern und die Adapter dementsprechend zu pflegen. Sowohl das Trace-Format OPEN.xtrace als auch die Adapter für ausgewählte APM-Werkzeuge sind auf GitHub quelloffen veröffentlicht (s. AP 6).

OPEN.xtrace wurde auf der 13th European Workshop on Performance Engineering (EPEW 2016) veröffentlicht und präsentiert (s. AP 6) [Ok16]. Es wurde ebenfalls ein Beitrag zu OPEN.xtrace auf dem 7. Symposium on Software Performance (SSP 2016) präsentiert. Die OPEN.xtrace-Initiative traf sowohl in Wissenschaft als auch Industrie auf sehr positives Interesse, da das OPEN.xtrace-Format über diagnoseIT hinaus nutzbar ist — etwa für aktuell Werkzeug-spezifische Ansätze zur Extraktion von Performance-Modellen.

Es wurde beschlossen, dass OPEN.xtrace zukünftig im Rahmen der Research Group der Standard Performance Evaluation Corporation (SPEC RG) weiterentwickelt und veröffentlicht wird. Hiervon versprechen wir uns eine höhere Sichtbarkeit und Akzeptanz. Hierzu wurde eine Webpräsenz für APM-Interoperabilität unter `http://research.spec.org/apm-interoperability` eingerichtet, unter der OPEN.xtrace und dessen Weiterentwicklung verfügbar sein wird. Langfristig versuchen wir, OPEN.xtrace als Austauschformat in der APM-Community zu etablieren.

Der OPEN.xtrace-Beitrag ist ein Kern, der in diagnoseIT angestrebten APM-Werkzeugunabhängkeit.

### 3.2   AP 2: APM-Wissensbasis

Ziel des Arbeitspakets war die Entwicklung der Infrastruktur für die APM-Wissensbasis sowie deren Befüllung mit APM-Expertenwissen.

Eine Entscheidung, die sehr früh im Projektverlauf getroffen wurde, ist, dass zunächst ein regelbasierter Ansatz entwickelt und evaluiert werden soll. Die Kernidee ist, dass für die Diagnose ein Satz Regeln existiert, wobei jede Regel einen Erkenntnisgewinn und ggf. eine Verfeinerung der Instrumentierung für weitere Regeln generiert [He16]. Prototypische Diagnoseregeln für typische Performance-Probleme wurden definiert. Diese analysieren Traces, die mittels OPEN.xtrace (AP 1) ausgedrückt sind, sowie Zeitreihen aufgezeichneter Systemmetriken wie CPU-Auslastung. Als Sprache für die Formalisierung der Regeln hatten wir uns zunächst für Drools entschieden. Später haben wir eine eigene *diagnoseIT*-Rule-Engine entwickelt.

Der Regelsatz basiert auf aus der Literatur und der APM-Praxis bekannten Performance-Anti-Patterns. Es wird zwischen generischen und semantifizierenden Regeln unterschieden. Die generischen Regeln können immer angewendet werden. Die semantifizierenden Regeln bilden technologie- und systemspezifisches Wissen auf Regeln ab und basieren unter anderem auf generischen Performance-Anti-Patterns. Der Erkennungsansatz lässt sich anhand der verwendeten Datenbasis aufteilen in die folgenden drei Kategorien: Trace-basiert, Zeitreihen-basiert und eine Kombination aus den beiden zuvor genannten Kategorien. Dabei wurden für die Kategorie Trace-basiert fünf Regeln (N+1, Stifle, Circuitous Treasure Hunt, Expensive Computation, Phantom Logging), für die Kategorie Zeitreihen-basiert drei Regeln (Ramp, Traffic Jam, More is Less) und für die Kombination aus beiden zwei Regeln (Application Hiccups, Garbage Collection Hiccups) erstellt. Der Auswahl der Performance-Antipatterns ging eine Analyse voraus, inwieweit sich Antipatterns mit dem diagnoseIT-Ansatz erkennen lassen. Ferner wurde der Regelsatz für Szenarien mit Mobilgeräten und Ende-zu-Ende-Kommunikation erweitert [An17a].

Aus Diskussionen u. a. mit den assoziierten Partnern stellte sich der generische Teil des Regelsatzes bereits als gewinnbringend heraus. Es wurde angeregt, eine Möglichkeit zu schaffen, die gefundenen Probleminstanzen durch anwendungsspezifische Meta-Informationen (z.B. „Das Problem ist bekannt und soll in der nächsten Version behoben sein") anzureichern. Außerdem wurde angeregt, den Regelsatz um Trace-übergreifende Konzepte und Korrelation mit Systemmetriken (z.B. CPU- und Thread-Pool-Auslastung) zu erweitern. Beides wurde im Projekt aus Zeitgründen nicht mehr behandelt, stellt aber eine vielversprechende zukünftige Arbeit dar.

Durch den Austausch mit den assoziierten Partnern hat sich außerdem ergeben, dass eine sinnvolle Kategorisierung der Diagnoseergebnisse einen großen Mehrwert liefert. Basierend auf dieser Erkenntnis wurde eine automatisierten Kategorisierung von Diagnoseergebnissen entwickelt [An17b]. In diesem Kontext wurden unterschiedliche Clustering-Algorithmen hinsichtlich der Eignung für diesen Anwendungsfall untersucht und k-Means wie auch hierarchisches Clustering als geeignet ausgewählt. Die Idee ist es, einen selbstlernenden Algorithmus zu entwickeln, der die optimale Konfiguration des Clusterings selbstständig kalibriert und so eine möglichst aussagekräftige Kategorisierung der Diagnoseergebnisse bereitstellt.

Zur effizienten Bestimmung einer sinnvollen Instrumentierung in einem neuen Anwendungskontext von diagnoseIT haben wir die Ausarbeitung eines generischen Instrumentierungsprozesses begonnen. Der Instrumentierungsprozess soll Anwender mithilfe eines begleitenden Fragebogens zielgerichtet zu einer geeigneten Konfiguration der Instrumentierung für ihr Software-System führen. Diese Aktivität soll nach Projektabschluss fortgeführt werden.

Komplementär zum Projektfokus auf der Diagnose von APM-Daten wurde ein Ansatz entwickelt, um Performance-Antipatterns aus Profiler-Daten und Lasttests zu erkennen [Tr17].

### 3.3 AP 3: Laufzeitdiagnose

Ziel des Arbeitspakets war die Entwicklung von Laufzeitanalysen auf Basis der Systeminformationen (AP 1 und AP 4) sowie der APM-Wissensbasis (AP 2). Zu den Analysen zählen die Extraktion von Architekturinformation, die Erkennung und Diagnose von Performance-Problemen sowie Anpassung der Instrumentierung.

Initial wurde für das EPM (aus AP 1) eine Komponente zur Systemmodellwartung entwickelt, die ein EPM aus eingehenden Trace-Daten und Ressourcen-Messungen erstellt. Nach der Entscheidung die Analyse-Komponente möglichst zustandslos zu konzipieren, haben wir den Fokus der Analysekomponente auf die Diagnose von Performance-Problemen verlagert.

Die Analysekomponente zur Muster-Interpretation und Durchführung der Diagnose wurde zunächst im Wesentlichen durch die Drools-Laufzeitumgebung realisiert. Diese wurde im weiteren Projektverlauf durch eine leichtgewichtigere, speziell in Bezug auf die Anforderungen von *diagnoseIT* entwickelten, Rule-Engine ersetzt. Dabei soll die speziell für *diagnoseIT* entwickelte Rule-Engine die Flexibilität und einfache Erweiterbarkeit der Diagnoseregeln ermöglichen.

Durch eingehende Analysen der gängigen, freien und kommerziellen APM-Werkzeuge wurde die Entscheidung getroffen, die Funktionalität zur Symptomerkennung in die Verantwortlichkeit der APM-Werkzeuge zu legen. Welche Aufrufe (bzw. Traces) langsam oder „problembehaftet" sind, können die APM-Werkzeuge selbst am besten entscheiden, da diese in der Regel auch die hierfür notwendige Historie der Performance-Metriken vorhalten. Somit sieht *diagnoseIT* vor, dass die APM-Werkzeuge Traces als „problembehaftet" markieren und nur diese zur Diagnose an die diagnoseIT-Laufzeitumgebung propagieren.

Im Rahmen dieser Entscheidung wurden Verfahren zur Anomalieerkennung konzipiert und entwickelt, die der Symptomerkennung für diagnoseIT dienen. Hierbei wurden zwei Aspekte betrachtet. (i) Im Vergleich zu existierenden Ansätzen zur Anomalieerkennung betrachtet der diagnoseIT-Anomalieerkennungsansatz einzelne Anfragen bei der Berechnung einer Anomalie, anstatt aggregierte Daten zu betrachten. Auf diese Weise ist es möglich, einzelne Anfragen und damit die entsprechenden Traces als „problematisch" zu markieren und für die weitere Diagnose bereitzustellen. (ii) In modernen Systemumgebungen (z. B.

Microservice-Architekturen) sind Änderungen an der Architektur und Systemlandschaft an der Tagesordnung. Dies stellt für die Erkennung von Anomalien eine Herausforderung dar, da Baselines nicht zuverlässig aus der Historie gelernt werden können. Es wurde ein prototypischer Ansatz konzipiert, der es Anomalieerkennungsverfahren erlaubt, diesen Aspekt explizit zu berücksichtigen. Mit der Umsetzung der beiden Aspekte der Anomalieerkennung konnten zwei initiale Annahmen für diagnoseIT aufgehoben werden: Die Annahme, dass Traces bereits als „problematisch" markiert sind und die Annahme dass die Software-Architektur sich nicht zur Laufzeit ändert. Ferner wurde ein hierarchischer Ansatz zur Vorhersage von Performance-Anomalien, basierend auf einer Kombination auf Vorhersagemodulen für einzelne Komponenten (z.B. Zeitreihen, maschinelles Lernen) und Architekturinformation (u.a. zur Fehlerpropagation) entwickelt [Pi17, Pi17].

Die *diagnoseIT*-Analysekomponente wurde im Projektzeitraum an unterschiedliche freie und kommerzielle APM-Werkzeuge angebunden, sodass eine werkzeugunabhängige Erkennung von Mustern und Diagnose von Problemen gezeigt werden konnte. Neben der Anbindung der Analysekomponente an APM-Werkzeuge wurde eine prototypische SaaS-Version (Software as a Service) entwickelt, die Nutzern erlaubt, Traces in unterschiedlichen, werkzeugspezifischen Formaten für den Zweck der Diagnose hochzuladen. Die Idee der SaaS-Analysekompopnente bestand darin, unter Zurhilfenahme von OPEN.xtrace (AP 1) die hochgeladenen Traces in ein einheitliches Format zu transformieren und diese dann durch die Rule-Engine der Diagnose zu unterziehen. Im Laufe des Projekts wurde das Vorhaben nicht abgeschlossen. Zur Evaluierung der Idee, wurden im ersten Schritt die hochgeladenen Traces manuell analysiert.

### 3.4  AP 4: Werkzeug-Anbindung

Ziel des Arbeitspakets ist die Entwicklung der Schnittstelle zu den APM-Werkzeugen sowie die Integration und Erweiterung der Werkzeuge inspectIT und Kieker.

Für die Werkzeuge Kieker und inspectIT wurden Importer und Exporter entwickelt, die die jeweiligen Werkzeug-spezifischen Trace-Formate in das OPEN.xtrace-Format transformieren. Es wurde eine Sammlung von diagnoseIT-spezifischen Anforderungen an APM-Werkzeuge erstellt und deren Abgleich mit dem aktuellen Entwicklungsstand von inspectIT und Kieker durchgeführt. Im Projektverlauf wurden bisher nicht unterstützte Anforderungen in inspectIT und Kieker umgesetzt.

Für die APM-Werkzeuge inspectIT, Kieker, Dynatrace, CA APM, AppDynamics, New Relic, Riverbed und IBM APM wurde untersucht, wie sich Traces exportieren lassen und wie die exportierten Traces in das OPEN.xtrace-Format konvertiert werden können. Es wurden entsprechende Adapter entwickelt. Die Adapter sind als Open-Source-Software verfügbar und können von der Community genutzt werden.

In einem frühen Projektstadium wurde ein erster diagnoseIT-Prototyp realisiert, der die durch das Werkzeug gesammelten Traces an die Laufzeitkomponente weitergibt und die Analyseergebnisse in eine bestehende grafische Oberfläche einbindet. Dieser Prototyp war ganz bewusst zunächst nur in inspectIT integriert. Bei Tests zur Anbindung des Prototypen an die Systeme der assoziierten Partner wurden notwendige Erweiterungen zur Sicherstellung der Kompatibilität der Werkzeuge identifiziert. Hierbei handelt es sich z. B. um Erweiterungen bei proprietären Trace-Formaten mit dem OPEN.xtrace-Format und der Anbindung von Kieker. Einige Erweiterungen wurden im Rahmen des Projekts umgesetzt. Im Hinblick auf die geplante Korrelation mit Systemmetriken wurden weitere Messsensoren entwickelt (u. a. Auslastung von Festplatte, Netzwerk, . . . ).

### 3.5   AP 5: Evaluation von diagnoseIT

Ziel des Arbeitspakets ist die qualitative und quantitative Bewertung der in den Arbeitspaketen AP 1–4 erarbeiteten Ergebnisse. Die Evaluation erfolgte in Form von Labor- und Fallstudien.

**Laborstudien**   Zur qualitativen und quantitativen Evaluation in Laborstudien wurden das Netflix-RSS-Beispiel und der DVD Store sowie synthetische Benchmarks verwendet. In den synthetischen Benchmarks (z. B. zum Stresstest der Laufzeitumgebung sowie der Evaluation des Regelsatzes) werden neben Traces aus den Netflix- und DVD-Store-Beispielen auch Traces aus den Fallstudiensystemen (s.u.) verwendet. Der DVD-Store wird auch zur Demonstration des Prototypen verwendet. Das Netflix-Beispiel dient zur Evaluation des Monitorings in verteilten Systemen (u. a. wurden inspectIT und Kieker für das verteilte Monitoring mit diesen Technologien erweitert).

Im Rahmen einer Masterarbeit wurden qualitative und quantitative Laborstudien zur Evaluation des EPM (inkl. Repository) sowie der Laufzeitumgebung zum Einlesen von Traces und Ressourcenmessungen durchgeführt.

Für inspectIT und Kieker wurde eine bestehende Open-Source-Beispielanwendung (Netflix RSS) aufbereitet. Dies umfasst u. a. die Bereitstellung mittels der Container-Virtualisierungstechnologie Docker inkl. Varianten mit Instrumentierung für inspectIT und Kieker. Diese Ergebnisse sind bereits über die verbreiteten Plattformen GitHub/DockerHub verfügbar. Neben der Netflix-RSS-Anwendung wurde eine weitere Anwendung (SockShop) verwendet, die eine höhere Komplexität aufweist.

Im Kontext der SPEC RG wirken wir maßgeblich an einer Beispiel-Umgebung für „Performance-Aware DevOps" mit, in der Komponenten zum Monitoring (inkl. Aufzeichnung und Analyse von OPEN.xtrace-Traces) sowie der Performance-Problem-Erkennung, -Diagnose und -Vorhersage eine zentralen Rolle spielen [Du17]. Neben den konzeptionellen Arbeiten erfolgt eine Umsetzung auf Basis von RedHat fabric8 (`http://fabric8.io/`).

Für Laborexperimente wurde eine Bibliothek entwickelt, welche die gezielte Injektion von Performance-Problemen (Anti-Pattern) in Java EE-Anwendungen mittels Bytecode-Modifikation erlaubt [Ke16].

**Fallstudien bei den assoziierten Partnern** Die assoziierten Partner haben jeweils mit einem Fallstudiensystem zum Projekt beigetragen:

- System 1 ist eine zentrale Integrationsplattform zur Abwicklung von Zahlungsverkehr. Zu Projektbeginn waren keine Monitoring- und/oder Diagnose-Lösungen im Projekt im Einsatz.
- System 2 ist eine Java-Enterprise-Anwendung zur Abwicklung von Versicherungs-verträgen. Als APM-Werkzeug ist dynatrace im Einsatz. Die Konfiguration der Messdatengewinnung und die Diagnose erfolgt komplett manuell.

Das initiale Vorgehen zur Einbindung der beiden assoziierten Partner war analog: Es gab zunächst jeweils einen eintägigen Besuch der Projektpartner vor Ort, in dem der aktuelle Projektstand (inkl. Demonstration mit dem Prototypen) präsentiert und diskutiert wurden. Im Anschluss gab es eine strukturierte Diskussion, in der über konkrete Pläne zur Einbindung der Systeme diskutiert wurde. Von den assoziierten Partnern waren jeweils ca. zehn Vertreter aus Entwicklung und Betrieb bei den Treffen vertreten. Ergebnisse dieser Treffen waren Folgetermine auf Arbeitsebene. Bei den Folgeterminen wurden neue Herausforderungen für die Diagnose identifiziert (z. B. neue Problemklassen), die im weiteren Projektverlauf berücksichtigt werden.

Die Anbindung an diagnoseIT erfolgt jeweils über OPEN.xtrace. Bei System 1 wurde inspectIT integriert. Bei System 2 erfolgte die Anbindung über den entsprechenden Adapter. Aus beiden Systemen standen Traces für die Analyse zur Verfügung. Diese wurden zur Evaluation des Regelsatzes verwendet. In beiden Fällen konnten Anti-Patterns in den Traces erkannt werden (z. B. Phantom Logging). Die Evaluation diente ebenfalls als Test für die OPEN.xtrace-Adapter.

Im weiteren Projektverlauf haben weitere Treffen mit den assoziierten Partnern stattgefunden, um Informationen zur zukünftigen Ausgestaltung der Regeln zu sammeln und das weitere Vorgehen im Hinblick auf die Fallstudien abzustimmen. In diesem Zusammenhang wurde bei einem Partner eine neue Version des Prototyps eingerichtet.

**diagnoseIT-as-a-service** Um eine einfache Möglichkeit zu schaffen, Traces von weiteren externen Interessierten zu bekommen, wurde eine Webseite eingerichtet: http://ditaas.inspectit.rocks/. Wir erhoffen uns dadurch einen breiteren Satz an Traces (in Form der abgedeckten APM-Werkzeuge/Formate und Problemklassen) zu erhalten. Aktuell werden die eingereichten Traces gespeichert und offline in die Laufzeitumgebung eingespielt. Auf längere Sicht ist geplant, die Laufzeitumgebung direkt an den Dienst anzubinden, um eine automatisierte Analyse und Antwort umzusetzen.

### 3.6 AP 6: Projektkoordination und Dissemination

**Projektkoordination**   Dieser Teil des Arbeitspakets umfasst sämtliche Aktivitäten die mit der Organisation, Erfolgskontrolle, Steuerung und Berichterstattung (Zwischenberichte und Abschlussbericht) des Projektes verbunden sind.

Das Projekt wurde durch wöchentliche Telefonkonferenzen und monatliche Workshops geführt. Das Vorgehensmodell, das für die Projektabwicklung verwendet wurde, ist angelehnt an agile Software Entwicklungsmethoden wie beispielsweise Scrum und durch eine iterative Herangehensweise charakterisiert. Ähnlich zur Idee von Daily-Scrum-Meetings wurden in diesen Treffen der aktuelle Arbeitsfortschritt aufgezeigt, Probleme diskutiert und die Arbeitsplanung abgestimmt. An diesen Treffen nahmen in der Regel alle am Projekt beteiligten Personen von der NovaTec und der Universität Stuttgart teil. Die räumliche Nähe der Universität Stuttgart und der NovaTec Consulting GmbH ermöglichte eine enge und sehr gute Kooperation der Projektpartner untereinander, wodurch auch die Durchführung von kurzfristig organisierten Treffen möglich und besonders hervorzuheben ist.

Darüber hinaus erfolgt eine asynchrone Abstimmung der Projektbeteiligten über Plattformen wie Skype und Atlassian Confluence. Im Confluence-System existiert eine umfassende Dokumentation von Projektartefakten wie z. B. Protokolle aller Telefonkonferenzen und anderer Besprechungen (z. B. Workshops), Treffen mit assoziierten Partnern und Entwurfsdokumente. Zusätzlich wurde ein dedizierter geschützter Bereich in Confluence zur Zusammenarbeit mit externen Partnern genutzt, zu denen auch Studenten gehören, die Themen im diagnoseIT-Kontext bearbeitet haben.

Die Qualitätssicherung wurde durch das agile Vorgehensmodell sowie gemeinsam durchgeführte Reviews und Evaluationen in besonderem Maße unterstützt. Der Rahmen für die Qualitätssicherung war dabei durch die gemeinsamen Workshops gegeben.

Es wurden zudem interne diagnoseIT-Klausurtagungen durchgeführt, davon zwei im Schloss Dagstuhl, dem weltweit bekannten Begegnungszentrum für Informatik:

- 20.—23. März 2016, Event 16124, `http://www.dagstuhl.de/16124`
- 04.—07. Dezember 2016, Event 16494, `http://www.dagstuhl.de/16494`

Die Workshops wurden genutzt, um den bisherigen Projektverlauf zu reflektieren, einzelne Themen im Detail zu diskutieren und Pläne für den weiteren Projektverlauf zu entwickeln.

Die assoziierten Partner wurden konkret in das diagnoseIT-Projekt eingebunden. Der Fokus der Zusammenarbeit mit den assoziierten Partnern lag auf dem Einholen von Feedbacks sowie der Vorbereitung und Durchführung der Evaluation.

Für die Unterstützung der Projekttätigkeiten kamen für verschiedene Belange unterschiedliche Werkzeuge zum Einsatz. Zur Qualitätssicherung von Code-Artefakten wurde ein Continuous-Integration-System (Jenkins CI) verwendet. Code-Artefakte und Dokumente sowie Änderungen wurden mit Hilfe eines Code-und-Dokument-Repository (GitHub, GitLab) verwaltet. Für den asynchronen Austausch der Projektpartner und für die Wissensdokumentation kam ein Wiki (Atlassian Confluence) zum Einsatz sowie ein Issue/Tickettracking-System (Atlassian JIRA) für die Aufgabenplanung. Diese Systeme waren den Projektpartnern frei zugänglich, leichtgewichtig und bereits im industriellen Kontext für das Projektmanagement und zur Projektabwicklung etabliert. Die Werkzeuge wurden als Software-as-a-Service genutzt — mit Ausnahme des Continuous-Integration-Systems, das als Amazon EC2-Instanz von der NovaTec Consulting GmbH bereitgestellt wurde.

**Dissemination** Dieser Teil des Arbeitspakets umfasst sämtliche Aktivitäten, die dazu dienen, die in *diagnoseIT* erarbeiteten Forschungsergebnisse während der Projektdurchführung einer breiteren Öffentlichkeit zur Verfügung zu stellen. Hierzu zählen beispielsweise *i)* die Einrichtung und Pflege einer *diagnoseIT*-Webpräsenz, *ii)* die Veröffentlichung der Projektergebnisse auf Veranstaltungen wie Messen und Fachtagungen sowie in Fachzeitschriften, *iii)* die Veröffentlichung von entwickelter Software.

Für das Projekt wurde eine Webpräsenz unter `http://diagnoseit.github.io/` eingerichtet. Für das Projekt wurden außerdem Zugänge bei den Social-Media-Plattformen LinkedIn und Twitter eingerichtet.

Im Juni 2015 wurde eine Presseerklärung zum diagnoseIT-Projektstart erstellt. Diese wurde u. a. auf den diagnoseIT, NovaTec- und Kieker-Webseiten sowie LinkedIn, LinkedIn-Novatec, Twitter, XING-Novatec, Facebook-Novatec, Google+, OpenPR veröffentlicht. Ferner gab es einen Beitrag zum Projekt im SPEC RG Newsletter (vol. 2, issue 1, 2016).

Im Rahmen des Projekt wurden zahlreiche Publikationen in Zeitschriften, Konferenzen und Workshops erstellt. Es wurden zahlreiche Vorträge, Demos und Tutorials gehalten. Eine Vielzahl von Studierenden wurden in das Projekt eingebunden und haben zum Projekterfolg beigetragen. Eine detaillierte Auflistung der in diesem Absatz genannten Punkte ist Abschnitt 3.11 zu entnehmen.

Ferner waren wir während des Projektzeitraums in verschiedenen Rollen (Funktionen, Sponsoring etc.) an der Organisation von wissenschaftlichen Veranstaltungen beteiligt:

- ACM/SPEC International Conference on Performance Engineering ('15, '16, '17), `http://icpe-conference.org/`

- International Workshop on Quality-Aware DevOps ('15, '16, '17), `http://qudos-workshop.org/`

- Symposium on Software Performance ('15, '16, '17), `http://performance-symposium.org`

- Dagstuhl seminar on „Software Performance Engineering in the DevOps World", `http://www.dagstuhl.de/16394`

### 3.7 Wichtigste Positionen des zahlenmäßigen Nachweises

Die Kosten bzw. Ausgaben entfielen zum weitaus größten Teil auf die Position 0837 (Personalkosten) und zu einem sehr kleinen Teil auf die Position 0838 (Reisekosten). Auf die weiteren Positionen des Verwendungsnachweises entfielen keine Kosten bzw. Ausgaben.

### 3.8 Notwendigkeit und Angemessenheit der geleisteten Arbeit

Die automatische Problemerkennung und -diagnose stellen ein für die wirtschaftliche und wissenschaftliche Verwertung und den Anschluss herausragendes Alleinstellungsmerkal im Kontext des Application Performance Management dar. Gleichzeitig sind die notwendigen Projektarbeiten für ein KMU eine Forschungsinvestition mit nicht unerheblichem Risiko. Das Projekt *diagnoseIT* hat die Grundlage für die wirtschaftliche Umsetzung geschaffen und wesentliche wissenschaftliche Beiträge geleistet. Das Geschäftsfeld Application Performance Management der NovaTec Consulting GmbH ist nachhaltig gestärkt, es sind wissenschaftliche Folgeprojekte entstanden (u. a. ContinuITy), und das Projektkonsortium plant weitere gemeinsame Aktivitäten. Alle durchgeführten Projektarbeiten und Ergebnisse waren dazu notwendig und angemessen.

### 3.9 Darstellung des voraussichtlichen Nutzens

Wie im vorangegangenen Abschnitt erläutert, konnten während der Projektlaufzeit die erstellten Artefakte in Laborstudien und Fallstudien erfolgreich evaluiert werden. Zudem gab es bei der Präsentationen des Ansatzes vor Vertretern der Industrie und der Wissenschaft stets großes Interesse. Somit kann ein großes Marktpotential unterstellt werden.

### 3.10 Fortschritte bei anderen Stellen

Während des Projekts wurde in die kommerzielle APM-Lösung Dynatrace AppMon eine Erkennung von Problemmustern integriert (`https://www.dynatrace.com/blog/automatic-`

problem-detection-with-dynatrace/). Dieses Feature erlaubt es, eine Reihe von bekannten Problemmustern zu erkennen. Diese Muster werden in die verschiedenen Kategorien Antwortzeit, Komplexität, Threading, Asynchronous, Content Type, Database, Performance Breakdown, Log Activity, and HTTP Response Codes unterteilt. Pro Kategorie gibt es sogenannten Buckets die die Problemmuster genauer definieren. Jeder Nutzerrequest wird dann mit diesen Buckets getagt sobald eines dieser Muster erkannt wird. Somit kann auf die Nutzerrequests gefiltert werden, die ein bestimmtes Problemmuster haben. Im Gegensatz zum *diagnoseIT*-Projekt werden aber keine Root-Cause-Analysen durchgeführt, um genau die Methoden und Codestellen zu erkennen, die für die Performance-Probleme verantwortlich sind. Dieser Schritt muss bei Dynatrace AppMon manuell erfolgen.

Zudem wurde während des Projekts die OpenTracing API eingeführt. OpenTracing API ist ein API-Standard für das Tracing von verteilten Anwendungen. Es ermöglicht Entwicklern von Anwendungscode, ihren Code zu instrumentieren, ohne sich an einen bestimmten Trace-Anbieter zu binden. Hierbei wird jedoch das Format und die Datenstruktur der Tracing-Informationen nicht vorgegeben. Im Gegensatz dazu definiert OPEN.xtrace ein konkretes Format für den Austausch von Execution Traces zwischen APM-Werkzeugen.

### 3.11 Erfolgte und geplante Veröffentlichungen

Im Zusammenhang mit dem *diagnoseIT*-Projekt sind die im Folgenden aufgelisteten Publikationen, Vorträge und studentischen Arbeiten erfolgt. Bei den Publikationen ist zusätzlich eine Referenz auf das enthaltene Literaturverzeichnis enthalten.

#### 3.11.1 Publikationen

**Zeitschriftenartikel**

- [Pi17] Pitakrat, Teerat; Okanović, Dušan; van Hoorn, André; Grunske, Lars: Hora: Architecture-aware online failure prediction. Journal of Systems and Software, 2017. In press. Online first: `https://doi.org/10.1016/j.jss.2017.02.041`

- [Tr17] Trubiani, Catia; Bran, Alexander; van Hoorn, André; Avritzer, Alberto; Knoche, Holger: Exploiting Load Testing and Profiling for Performance Antipattern Detection. Information and Software Technology, 2017. In press. Online first: `https://doi.org/10.1016/j.infsof.2017.11.016`

Ferner ist eine Veröffentlichung, die den diagnoseIT-Ansatz im Detail beschreibt und evaluiert, in Erstellung. Diese wird zeitnah bei einer einschlägigen Zeitschrift eingereicht.

**Workshop- und Konferenzpublikationen**

**Technische Berichte**

- [HvH15] Hasselbring, Wilhelm; van Hoorn, André: Open-Source Software as Catalyzer for Technology Transfer: Kieker's Development and Lessons Learned. Bericht TR-1508, Department of Computer Science, Kiel University, Germany, August 2015

- [Br15] Brunnert, Andreas; van Hoorn, Andre; Willnecker, Felix; Danciu, Alexandru; Hasselbring, Wilhelm; Heger, Christoph; Herbst, Nikolas; Jamshidi, Pooyan; Jung, Reiner; von Kistowski, Joakim; Koziolek, Anne; Kroß, Johannes; Spinner, Simon; Vögele, Christian; Walter, Jürgen; Wert, Alexander: Performance-oriented DevOps: A Research Agenda. Bericht SPEC-RG-2015-01, SPEC Research Group — DevOps Performance Working Group, Standard Performance Evaluation Corporation (SPEC), August 2015

**Poster und Demonstrationen**

- van Hoorn, A., Siegl, S.: Application performance management (APM): Continuous monitoring of application performance (Fachposter im OBJEKTspektrum, auf Deutsch). https://www.sigs-datacom.de/wissen/fachposter/ (2017)

- Heger, C., van Hoorn, A., Okanović, D., Siegl, S., Wert, A.: diagnoseIT: Expert-guided Automatic Diagnosis of Performance Problems in Enterprise Applications (Poster und Demo). 7th ACM/SPEC International Conference on Performance Engineering (ICPE '16). **Best Demo Award.**

### 3.11.2 Vorträge und Tutorials

Neben den Vorträgen für die oben aufgeführten Publikationen auf Workshops und Konferenzen haben wir die folgenden Vorträge und Tutorials gehalten.

**Konferenz- und Workshopvorträge (ohne entsprechende Publikation)**

- Angerstein T., Hidiroglu A., Palenga M., Sassano M., Monitoring and Diagnosis of Performance Problems in Enterprise Applications With Mobile Front-end, Symposium on Software Performance 2017, Karlsruhe, Germany

- Bran, A., Hidiroglu A., Palenga , M., Okanović, D., APM Interoperability with OPEN.xtrace: Overview and Lessons learned, Symposium on Software Performance 2016, Kiel, Germany

- Heger C., van Hoorn, A., Okanović, D., Siegl, S., Wert , A., Fighting Groundhog Days: Expert-guided Automatic Diagnosis of Performance Problems in Enterprise Applications, Symposium on Software Performance 2015, Munich, Germany

**Tutorials**

- Heger, C., van Hoorn, A., Mann, M., Okanović, D.: Application performance management: State of the art and challenges for the future. 3-stündiges Tutorial auf der 8th ACM/SPEC International Conference on Performance Engineering (ICPE '17). Verbunden mit der Publikation [He17].

**Vorträge auf Meetups (Auswahl)**

- Because performance matters! — Open Source Application Performance Monitoring for the Crowd, Presented in Frankfurt, Stuttgart, Munich, Kaiserslautern, Görlitz
- van Hoorn, A. Tool-supported application performance problem detection and diagnosis, Software Performance Meetup 2015, München

**Eingeladenen Vorträge an Universitäten**

- Siegl, S. Application Performance Management. 2016, 2017, University of Stuttgart.
- van Hoorn, A. Application Performance Management (APM), University of Ulm, 2016
- van Hoorn, A. Evaluating Software Performance with Models or with Measurements? Both—and Continuously! National University of Central Buenos Aires, Tandil, Argentinien, 2016
- van Hoorn, A. Measurement-Based Application Performance Problem Detection and Diagnosis, Gran Sasso Science Institute (GSSI), L'Aquila, Italy, 2015

### 3.11.3 Studentische Arbeiten

Die im Folgenden aufgeführten Arbeiten wurden in der Regel gemeinsam von der Universität Stuttgart und der NovaTec betreut.

**Masterarbeiten**

- Vogel, S. Automated Root Cause Isolation in Performance Regression Testing, University of Stuttgart, 2017.
- Düllmann, T. Change-Aware Performance Anomaly Detection in Microservice Architectures, Master's Thesis, University of Stuttgart, 2017.
- Oehler, M. Anomaly Detection Based on Continuous Monitoring with inspectIT, Master's Thesis, Hochschule der Medien and NovaTec, 2016.
- Waldvogel, C., Specification and Runtime Extraction of Enterprise Application Architectures for Expert-Guided Performance Problem Diagnosis, University of Stuttgart, 2015.

**Masterprojekte**

- Angerstein, T., Hidiroglu, A., Palenga, M., Spectrum-based Performance Problem Localization in Microservice Architectures. Angeleitete Forschung, University of Stuttgart, 2017.
- Angerbauer, K., Angerstein, T., Hidiroglu A., Lehmann S., Palenga, M., Röhrdanz, O., Sassano, M., Völker, C. Mobile-aware Diagnosis of Performance Problems in Enterprise Applications, Entwicklungsprojekt, 2017.
- Studien zu Forschungsmethoden der Softwaretechnik:
    - How well does diagnoseIT detect anti-patterns in real-world execution traces?
    - Does diagnoseIT help developers to diagnose performance problems?
    - Does architectural information from traces improve quality of performance problem clustering?

**Bachelorarbeiten**

- Sassano M., Evaluating Mobile Monitoring Strategies for Native iOS Applications, University of Stuttgart, 2017.
- Seifermann, V. Application Performance Monitoring in Microservice Based Systems, University of Stuttgart, 2017. In Zusammenarbeit mit Audi.
- Bran, A., Detecting Performance Anti-Patterns from Profiler Data, University of Stuttgart, 2017.
- Kunz, J. A Generic Platform for Transforming Monitoring Data into Performance Models, Karlsruhe Institute of Technology and NovaTec, 2016.

- Angerstein, T. Automated Categorization of Performance Problem Diagnosis Results, University of Stuttgart and NovaTec, 2016.
- Hidiroglu, A. Rule-Based Performance Anti-Pattern Detection from Execution Traces, University of Stuttgart and NovaTec, 2016.

**Bachelorprojekte**

- Bran, A., Hidiroglu, A., Palenga, M. Assesing the Interoperability of APM Trace Formats, Fachstudie, University of Stuttgart, 2016
- Benz, J., Sassano, M., Seifermann, V. Evaluation of Application Performance Management Tools in the Context of an Existing Enterprise IT Landscape. Fachstudie, University of Stuttgart, 2016. In Zusammenarbeit mit Volkswagen AG, Audi, AppDynamics, CA Technologies.

**Software**

- diagnoseIT: https://github.com/diagnoseIT
- OPEN.xtrace: https://github.com/spec-rgdevops/OPEN.xtrace
- diagnoseIT GUI prototype: http://bit.ly/2vi804N

# Literatur